\documentclass[showpacs,superscriptaddress]{revtex4}
\newcommand{\beq}{\begin{equation}}
\newcommand{\eeq}{\end{equation}}
\newcommand{\barr}{\begin{eqnarray}}
\newcommand{\earr}{\end{eqnarray}}

\usepackage{graphicx}
\usepackage{amsmath}
\usepackage{amssymb}

\begin{document}


\title{Generalization of the Hellmann-Feynman theorem}

\author{J. G. Esteve}
\affiliation{Departamento de F\'{\i}sica Te\'{o}rica, Facultad de Ciencias,
 Universidad de Zaragoza, 50009 Zaragoza, Spain}
\affiliation{Instituto de Biocomputaci\'on y F\'{\i}sica de Sistemas complejos 
(BIFI),
  Universidad de Zaragoza, 50009 Zaragoza, Spain}
\author{Fernando Falceto}
\affiliation{Departamento de F\'{\i}sica Te\'{o}rica, Facultad de Ciencias,
 Universidad de Zaragoza, 50009 Zaragoza, Spain}
\affiliation{Instituto de Biocomputaci\'on y F\'{\i}sica de Sistemas complejos 
(BIFI),
  Universidad de Zaragoza, 50009 Zaragoza, Spain}
 \author{C. Garc\'{\i}a Canal}
\affiliation{Laboratorio de F\'{\i}sica Te\'{o}rica,
Departamento de F\'{\i}sica, Facultad de Ciencias Exactas,
  Universidad Nacional de La Plata and IFLP-CONICET, Argentina}

\begin{abstract}
The well known Hellmann-Feynman theorem of Quantum Mechanics connected
with the derivative of the eigenvalues with respect to a parameter upon which 
the Hamiltonian depends,
is generalized to include
cases in which the domain of definition of the Hamiltonian of the system also 
depends
on that parameter.
\end{abstract}

\maketitle


\section{Introduction}

The Hellmann-Feynman theorem (from now on the HF theorem) is a useful tool in 
solid state, atomic and molecular physics. One of its consequences
being that in Quantum Mechanics there is a single way of defining 
a generalized force on eigenstates of the Hamiltonian,
associated to the variation of some of its parameters. 

Classically, given a Hamiltonian $U(\lambda)$ depending on a parameter 
$\lambda$ (which can be a generalized coordinate), one can define a 
generalized force $F_\lambda=-{\partial_\lambda}U$ 
which is associated with that parameter in the sense that 
$F_\lambda d\lambda$ is the work done in changing the parameter by 
$d\lambda$. 
However, in Quantum Mechanics there are, in principle, 
two possible ways to implement that definition: given the quantum 
Hamiltonian $H(\lambda)$ (with eigenvalues $E_n(\lambda)$ and normalized 
eigenvectors $\Psi_n(\lambda)$), we can define the generalized force acting 
on the state $\Psi_n(\lambda)$ as 
$ F_\lambda= -{\partial_\lambda E_n(\lambda)}$ or 
as the expectation value of $-{\partial_\lambda H(\lambda)}$ on the 
state $\Psi_n(\lambda)$, which gives 
$F_\lambda={\langle}\Psi_n|-{\partial_\lambda H(\lambda)}\Psi_n{\rangle}$. 
The HF theorem ensures that both definitions are equivalent, i. e.
\begin{equation}
\frac{\partial E_n (\lambda)} { \partial \lambda} = 
{\langle}\Psi_n(\lambda)|\frac{\partial H(\lambda)}
{ \partial \lambda}\Psi_n(\lambda){\rangle},
\label{uno}
\end{equation}
where, obviously, the differentiability of $E_n, H$ and $\Psi_n$
with respect to $\lambda$ is assumed. This equation
is known as the differential form of the HF theorem and from it 
we can obtain the integrated version,
\begin{equation}
E_n(\lambda_1) -  E_n(\lambda_2)  = 
\frac{{\langle}\Psi_n(\lambda_2)|(H(\lambda_1) -
  H(\lambda_2))\Psi_n(\lambda_1){\rangle}}{  
{\langle}\Psi_n(\lambda_2)|\Psi_n(\lambda_1){\rangle}},
\label{dos}
\end{equation}
and the off diagonal form,
\begin{equation}
(E_m(\lambda)-E_n(\lambda)){\langle}\Psi_n(\lambda)|\frac{\partial}{\partial 
\lambda} \Psi_m(\lambda){\rangle}=
{\langle}\Psi_n(\lambda)|\frac{\partial H}{\partial \lambda}
\Psi_m(\lambda){\rangle}.
\label{dosp}
\end{equation}
The equations (\ref{uno}-\ref{dosp}), and others derived from them have been 
used in many areas of physics and specially in solid state and molecular 
physics, with the pioneering work of Feynman \cite{F1} proving the, 
so called, electrostatic theorem. Taking $\lambda$ to be the coordinate 
$X_i$ of the position of the nucleus $i$ and assuming that there are no  
external fields, the HF theorem 
states that the force on nucleus $i$ due to the other nuclei and the 
electrons, in some particular configuration, is exactly what could be 
computed in classical electrostatics from the location of the other nuclei 
and the electronic charge density.
 Furthermore, the use of the HF theorem has been extended to variational 
states that are not
 the true eigenstates of the Hamiltonian or even to Gamow states 
(\cite{varia} - \cite{Gamow}).

However, there are also systems for which the HF theorem fails for the 
eigenstates of the Hamiltonian, even if they fulfill the general conditions 
established for the validity of the theorem (differentiability with respect 
to the parameter). As we shall see later, these situations are related to 
the fact that in quantum systems the Hamiltonian can depend on a parameter, 
not only through an explicit dependence in the operator
but also because the domain of definition of the Hamiltonian 
depends on that parameter. This happens, for instance, when the Hamiltonian 
contains interactions that are gauge invariant and then, the functional 
form of $H$ and the boundary conditions for the functions of its domain both 
depend on the particular gauge we chose. In this case, as we shall see, 
the standard derivation of the HF theorem is not valid and a generalization 
is needed to cover those systems. This is precisely the main 
goal of this contribution.

The article is organized as follows: In the next section we present a simple 
example of a Hamiltonian with magnetic interactions for which the HF theorem 
fails, in Section 3 we derive the generalization of the HF theorem,
the application to some Hamiltonians, including the one 
mentioned above, is contained in Section 4, 
and finally we sketch the conclusions.

\section{Particle interacting with a confined magnetic field}

\begin{figure}[h]
  \centering
   \includegraphics[width=4cm]{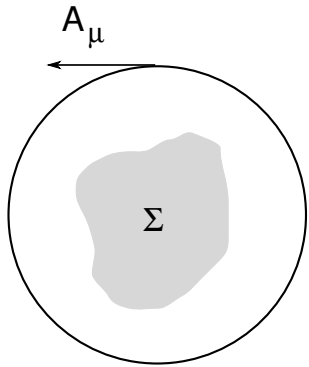}
  \caption{The planar rotor under the influence of a magnetic field confined in 
the region $\Sigma$}\label{rot}
\end{figure}

In this section we present an example of a system for which the standard form 
of the HF theorem (\ref{uno}) fails. The system describes a nonrelativistic 
charged particle moving on a circumference $S^1$ and interacting with a 
magnetic field which is confined in some region $\Sigma$ on the interior of 
$S^1$ with magnetic flux equal to $2 \pi \epsilon$ (see Fig. \ref{rot}). 
This is a model for the superconducting Cooper pairs moving on a ring that is 
transversed by a  magnetic flux $\Phi = 2 \pi \epsilon$ \cite{rotor}.

 Using suitable units,  and in a gauge where
 the electromagnetic field $A_\mu$ is tangent to the circumference, with
 constant norm $\epsilon$ and directed in the anticlockwise direction 
(consequently the flux through $\Sigma$ is
$\Phi=\int_\Sigma \vec{B} d \vec{S} = \int_{S^1} \vec{A} d\vec{l}=2 \pi\epsilon$), 
the Hamiltonian can be written
\begin{equation}
H=-\frac{1}{2} \left( \frac{d}{d \theta} - i \epsilon\right)^2,
\label{tres}
\end{equation}
acting on  periodic functions. The domain of definition of the Hamiltonian 
$D_0$ may be taken as the set of twice differentiable functions 
$f(\theta)$ (with $\theta \in [0,2 \pi]$), such that $f(0)=f(2\pi)$ and 
${\partial_\theta f}(0)={\partial_\theta f}(2\pi)$. 
In $D_0$, $H$ is essentially selfadjoint and has the eigenvalues 
and normalized eigenvectors:
\begin{eqnarray}
E_n(\epsilon)=\frac{1}{2} (n-\epsilon)^2 \\
\Psi_n(\theta)=\frac{1}{ \sqrt{2 \pi}} \exp{(in\theta)}.
\label {cuatro}
\end{eqnarray}
Note that the eigenvectors of the Hamiltonian do not depend on $\epsilon$. 
In this gauge, 
$\partial_\epsilon E_n(\epsilon)=\epsilon-n$,  
$\partial_\epsilon H(\epsilon)= i{\partial_\theta} + \epsilon$ and the HF equation (\ref{uno}) is exactly satisfied
\begin{equation}
\frac{\partial E_n(\epsilon)}{\partial \epsilon} = \epsilon -n = 
{\langle}\Psi_n|\frac{\partial H(\epsilon)}{\partial 
\epsilon}\Psi_n{\rangle}.
\label{r1}
\end{equation}

However, we could also work in another gauge obtained by the unitary 
 transformation
\begin{eqnarray}
H_g &= &e^{-i\epsilon \theta}\, H\,  e^{i\epsilon \theta} = -\frac{1}{2} 
\frac{d^2}{d \theta^2}
\\
D_{\epsilon}&= &e^{-i\epsilon \theta} \,D_0,
\label{Hg} 
\end{eqnarray}
so that the domain of definition $D_{\epsilon}$ of the transformed Hamiltonian is 
the set of  twice differentiable functions $f(\theta)$ (with $\theta \in [0,2 
\pi]$), such that 
\begin{eqnarray}
f(0) &=& e^{ i 2 \pi\epsilon}\, f(2\pi)\\
\frac {d f}{d \theta} (0) &=& e^{ i 2 \pi\epsilon}\, \frac{d f}{ d \theta} (2 
\pi).
\label{bc}
\end{eqnarray}

In $D_{\epsilon}$ the Hamiltonian $H_g$ is again essentially selfadjoint and, 
obviously, has the same  eigenvalues  that $H$ with eigenvectors
\begin{equation}
 \Psi_{g,n}(\epsilon,\theta)= \frac{1}{ \sqrt{2 \pi}} e^{i(n-\epsilon)\theta}.
\label {nn}
\end{equation}

In this gauge, the kinetic part of the Hamiltonian $H_{g}$ does not depend on 
$\epsilon$
 whereas its eigenvectors and the domain of definition  depend and, 
consequently,
 it is clear that the HF equation (\ref{uno}) is not valid for $H_g$ since
\begin{equation}
 {\langle} \Psi_{g,n}| \frac{\partial H_g}{\partial \epsilon} 
\Psi_{g,n}(\epsilon,\theta){\rangle}=0.
\end{equation}
 Now, the  HF theorem fails because with the gauge transformation we have 
transferred the dependence
 on the electromagnetic field from the kinetic part of $H$ to the domain of 
definition of $H_g$,
 a situation that is not taken into account in this form of the theorem.
 The key point here is that, because $D_{\epsilon}$ depends on $\epsilon$, 
some quantities like
\begin{equation}
\lim_{t\to 0} {\langle}\Psi_{g,n}|\frac{H_g(\epsilon+t)-H_g(\epsilon)}{t}| 
\Psi_{g,n}{\rangle}
\end{equation}
must be handled carefully because the domains of definition of $H_g(\epsilon + 
t)$ and $H_g(\epsilon)$ are different and $H_g$ is a non bounded operator.


\section{ Generalization of the Hellmann-Feynman theorem}\label{sec2}

In the previous section, we have seen an example that illustrates
the need of a generalization of HF theorem in order to include
the cases in which the domain of definition of the 
Hamiltonian also depends on the 
parameter we differentiate with respect to. 

Consider a selfadjoint Hamiltonian 
$H(\lambda)$ with domain ${\cal D}_\lambda$, 
eigenvalues $E_n(\lambda)$ and 
normalized eigenvectors $\Psi_n(\lambda)$. Assume
differentiability with respect to the parameter 
$\lambda$. 

Taking derivatives in the identity
\begin{equation}
 E_n(\lambda)={\langle}\Psi_n(\lambda)|H(\lambda) \Psi_n(\lambda){\rangle}
\label{id1}
\end{equation}
we find
\begin{eqnarray}
\frac{\partial E_n}{\partial \lambda} = {\langle}\frac{\partial 
\Psi_n}{\partial \lambda}|H\Psi_n{\rangle}
+ {\langle}\Psi_n| \frac{\partial H}{\partial 
\lambda}\Psi_n{\rangle}+
{\langle}\Psi_n|H\frac{\partial \Psi_n}{\partial\lambda}{\rangle}.
\nonumber
\end{eqnarray}

Note that in order to make sense of the previous expression, one has to extend 
$H(\lambda)$ to an operator defined in a domain that includes 
${\cal D}_\lambda$ for any $\lambda$.
In this case, the extended Hamiltonian will not be selfadjoint in general and 
therefore the expressions that involve such a non hermitian operator must be 
handled with care. That is the key point of the rest of the paper.

Adding and subtracting a term 
${\langle}H\Psi_n|
{\partial_\lambda}\Psi_n{\rangle} = 
E_n{\langle}\Psi_n|{\partial_\lambda} 
\Psi_n{\rangle} $, 
and using the fact that 
\begin{equation}
{\langle}\frac{\partial}{\partial \lambda} 
\Psi_n(\lambda)|\Psi_n(\lambda){\rangle}+{\langle}\Psi_n(\lambda)|
\frac{\partial}{\partial \lambda}\Psi_n(\lambda){\rangle}= 
\frac{\partial}{\partial \lambda} 
{\langle}\Psi_n(\lambda)|\Psi_n(\lambda){\rangle}=0
\end{equation}
for normalized eigenvectors, we finally obtain the 
desired generalization of the Hellmann Feynman theorem:
\begin{equation}
\frac{\partial E_n (\lambda)} { \partial \lambda}= 
{\langle}\Psi_n(\lambda)|\frac{\partial H(\lambda)}{ \partial 
\lambda}|\Psi_n(\lambda){\rangle} + \Delta_n (\lambda)
\label{gHF}
\end{equation}
with
\begin{equation}
\Delta_n (\lambda)={\langle} 
\Psi_n(\lambda)|H(\lambda)\frac{\partial }{\partial 
\lambda}\Psi_n(\lambda){\rangle}-{\langle}H(\lambda) \Psi_n(\lambda)|\frac{\partial }{\partial 
\lambda}\Psi_n(\lambda){\rangle}.
\label{gHF1}
\end{equation}
Compared with the equation (\ref{uno}), we have a new term 
$\Delta_n$ that vanishes for states $\Psi_n(\lambda)$ in ${\cal D}_\lambda$  
such that  
${\partial_\lambda}\Psi_n(\lambda)$ is also  in 
${\cal D}_\lambda$.
In this case
the normal form of HF theorem (\ref{uno}) is recovered. 
However, in general,
for states $\Psi_n(\lambda)$
such that  ${\partial_\lambda}\Psi_n(\lambda)\not\in{\cal D}_\lambda$,
the new factor $\Delta_n(\lambda)$ can give an extra contribution to the 
second term of the HF equation (\ref{gHF}).

The generalization of the other forms of the HF theorem can be obtained in a 
similar way. For the integrated form we have
\begin{eqnarray}
(E_n(\lambda_1) -  E_n(\lambda_2)) 
{\langle}\Psi_n(\lambda_2)|\Psi_n(\lambda_1){\rangle} &=& 
{\langle}\Psi_n(\lambda_2)|(H(\lambda_1) -  
H(\lambda_2)) \Psi_n(\lambda_1){\rangle}\nonumber \\
&+&\Delta_n(\lambda_1,\lambda_2),
\label{gdos}
\end{eqnarray}
with 
\begin{equation}
\Delta_n(\lambda_1,\lambda_2)=
{\langle}\Psi_n(\lambda_2)|H(\lambda_2)  \Psi_n(\lambda_1){\rangle}-{\langle}H(\lambda_2)\Psi_n(\lambda_2)|  
 \Psi_n(\lambda_1){\rangle}.
\end{equation}
And the off diagonal formulation is
\begin{equation}
(E_m(\lambda)-E_n(\lambda)){\langle}\Psi_n(\lambda)|\frac{\partial}{\partial 
\lambda} \Psi_m(\lambda){\rangle}=
{\langle}\Psi_n(\lambda)|\frac{\partial H(\lambda)}{\partial \lambda} 
\Psi_m(\lambda){\rangle} + \,\Delta_{n,m}(\lambda) 
\end{equation}
where the anomalous term is now
\begin{equation}
\Delta_{n,m}(\lambda)={\langle}\Psi_n(\lambda)|
H(\lambda) \frac{\partial \Psi_m(\lambda)}{\partial \lambda}{\rangle}
-{\langle}H(\lambda)\Psi_n(\lambda)|
 \frac{\partial \Psi_m(\lambda)}{\partial \lambda}{\rangle}.
\end{equation}

\section{Applications.}\label{sec4}

Now we can apply the equations (\ref{gHF},\ref{gHF1}) to the charged
planar rotor interacting with a confined magnetic field that was analyzed in 
section 2. For the Hamiltonian $H$ defined in (\ref{tres}) 
we obtain 
$\Delta_n(\epsilon)=0$ (because in this case $\partial_\epsilon\Psi_n \in  D_0 $)
 and we recover the equation (\ref{r1}). However, 
for $H_g$ we have:
\begin{equation}
\frac{\partial}{\partial \epsilon} \Psi_{g,n}(\epsilon,\theta) = -i \theta\, 
\Psi_{g,n}(\epsilon,\theta). 
\end{equation}
Consequently ${\partial_\epsilon} \Psi_{g,n}(\epsilon,\theta)$ 
is not in $D_{\epsilon}$. Now we can evaluate $\Delta_n$ as:
\begin{equation}
\Delta_n= {\langle}\Psi_{g,n}(\epsilon,\theta)\,|\, H_g 
\frac{\partial}{\partial \epsilon}\Psi_{g,n}(\epsilon,\theta){\rangle} \, -\,
{\langle} H_g \Psi_{g,n}(\epsilon,\theta)\,|\, \frac{\partial}{\partial 
\epsilon}\Psi_{g,n}(\epsilon,\theta){\rangle},
\end{equation}
that integrating by parts (or evaluated directly) gives a boundary term:
\begin{eqnarray}
\Delta_n &=& \frac{1}{2} \left\{ \left[ \frac{\partial}{\partial 
\theta}\Psi_{g,n}^*(\epsilon,\theta) 
\frac{\partial}{\partial \epsilon}\Psi_{g,n}(\epsilon,\theta) \right]_0^{2 
\pi}  \, -\,
 \left[  \Psi_{g,n}^*(\epsilon,\theta) \frac{\partial}{\partial \theta} 
\frac{\partial}{\partial \epsilon}\Psi_{g,n}(\epsilon,\theta) \right]_0^{2 
\pi} \right\}\nonumber\\
 &=& \epsilon -n.
\end{eqnarray}

It is interesting to note that in this case, the origin of the extra term 
$\Delta_n$ 
can be related to the anomalous Virial theorem for 
these kind of systems \cite{julio}.

Another example of a system where the standard form (\ref{uno})  of the HF 
theorem fails is a non relativistic particle in two dimensions
interacting with a $\delta(r)/r$ potential. This system has been 
studied in \cite{Jackiw,e1} as an example of anomalous symmetry breaking 
in quantum mechanics. The Hamiltonian
\begin{equation}
H=-\frac{\hbar^2}{2 m} \nabla ^2 + \lambda \delta^2(\mathbf{r})
\label{deltap}
\end{equation}
transforms under dilations, $\mathbf{r}\mapsto \alpha \mathbf{r}$, in a homogeneous way
$ H\mapsto \alpha^{-2} H $. This 
means that it can not have any normalizable eigenvector 
with energy different from zero, 
since if $\Psi_n(\mathbf{r})$ is an eigenstate with energy $E_n$, then 
$\Psi_n(\alpha \mathbf{r})$ is also an eigenstate with energy ${\alpha^{-2}} E_n$ 
for any real number $\alpha$. If $H$ is selfadjoint the previous
property implies that the only point in the discrete spectrum is
$E=0$. 

It is also known, however, that in  this system the $SO(2,1)$ 
symmetry is anomalously broken and there is  one s-wave normalized eigenstate 
\begin{equation}\label{eigen}
 \Psi_0(\alpha, \mathbf{r})= \frac {\alpha}{\pi^{1/2}} K_0 (\alpha r),
\end{equation}
which has an energy
 \begin{equation} 
E_0= -\frac{\hbar^2}{2 m} \alpha^2,
\end{equation}
 where in (\ref{eigen}) the $K_0$ Bessel function was used
 and the value of $\alpha$ determines the domain of the Hamiltonian.
The apparent contradiction is solved by the fact that although the 
Hamiltonian transforms mutiplicatively by a dilation its domain is not 
preserved and therefore the dilation does not actually produce
any new eigenvectors for the original Hamiltonian.

To see that, we can solve the s-wave sector of the Hamiltonian (\ref{deltap}) 
 by renormalizing the coupling $\lambda$ as in \cite{Jackiw}
 or by working in 
$ \mathbb{R}^2\setminus\{(0,0)\}$,
 in order to avoid the singularity at the origin
\cite{e1,albe}. In polar coordinates with Hamiltonian 
\begin{equation}\label{hd}
H_s=-\frac{\hbar^2}{2m} \left( 	\frac{\partial ^2}{\partial r^2} + 
\frac{1}{r} \frac{\partial}{\partial r} \right)\,,
\end{equation}
defined on the domain
\begin{equation}
D_{0}^\beta=\{ f\in L^2(R_+,rdr)| f\in C^\infty(R_+), c_0=\beta c_1\}
\end{equation}
with
\begin{eqnarray}
c_0&=&\lim_{r\rightarrow 0} \left(\frac{f(r)}{\log (\alpha_0 r)}\right)\\
c_1&=&\lim_{r\rightarrow 0}\left[ f(r)-c_0\log(\alpha_0 r)\right].
\label{dh}
\end{eqnarray}
In this domain $H_s$ is essentially selfadjoint. 
The meaning of this conditions is that if
$f(r)\in D_{0}^\beta$ then for $r\rightarrow 0;\,\, f(r)\sim
a(\log(\alpha r)+b)+o(1)$ with
\begin{equation}
\frac{1}{\beta}= b +\log\left(\frac{\alpha}{\alpha_0}\right)\,.
\label{14}
\end{equation}
and $\alpha_0$ the subtraction point. Now, it is easy to see that $
{\partial_\alpha}\Psi_0(\alpha,\mathbf{ r})=  (2/\alpha) \, G\,  \Psi_0(\alpha, \mathbf{r})$ where $G=\frac{1}{4}(\mathbf{ 
x}\mathbf{p}+\mathbf{p}\mathbf{x})$ is the infinitesimal generator of the dilations symmetry which is 
anomalously broken  because if $\Psi_0(\alpha, \mathbf{r}) \in D_0^\beta$ then
 $G\,\Psi_0(\alpha, \mathbf{r}) \in D_0^{\beta'}$ with $\beta' =\frac{\beta}{\beta +1}$.
  As the differentiation with respect to the parameter $\alpha$ do not preserves the domain, only the
general form of the HF theorem is valid.

In fact one can compute
\begin{eqnarray}
\frac{d E_0}{d\alpha}&=&  - \frac{\hbar^2}{m} \alpha\\
\frac{d H_s}{d\alpha}&=& 0 \\
 \Delta_0 &=& \frac{\hbar^2 \alpha}{m} 
\left\{ 
\left[ r \frac{d}{d r} K_0(\alpha r)
(r \frac{d}{d r} +1)K_0(\alpha r)
- r  K_0(\alpha r)\frac{d} {d r}
(r \frac{d}{d r} +1)K_0(\alpha r)\right]_0^\infty
\right\}\nonumber\\
&=&-\frac{\hbar^2 \alpha}{ m}\,.
\label{19} 
\end{eqnarray}
and the generalized Hellman-Feynman theorem (\ref{gHF}) holds. Notice that the evaluation of $\Delta_0$ do not requires the knowledge of $\Psi_0$ but only its behaviour at $r=0$ and $r=\infty$, which is fixed by the boundary conditions on the domain $D_0^\beta$.

There is a similarity between this system and the one analyzed previously,
 here we have eliminated the divergent term of the potential by avoiding the 
origin and 
introducing some boundary conditions for the functions of the domain of the 
Hamiltonian in that point and, in this way, 
we have transferred the dependence with the parameter from the Hamiltonian to 
its domain of definition, exactly the same that happened in the first example. 
On the other hand, the contribution of the $\Delta_0$ term 
is related to the property that
$G$ does not keep the domain of $H$ invariant,
 a fact whose main consequence is the appearance of a conformal 
anomaly in the system.

\section{Conclusion}

The main objective of this paper is to present a generalization of the 
well known Hellmann-Feynman theorem to enlarge its range of applicability.
In particular, our generalization allows to deal with
several interesting cases where the Hamiltonian of the system depends on 
a parameter, not only in the explicit expression of the operator, but also 
in its domain of definition.
We also present some remarkable examples to show a few applications 
of the generalization of the theorem. 


 This work is partially supported by
the grants FIS2006-01225   and FPA2006-02315, MEC (Spain) .
C.G.C. was partially supported by ANPCyT and CONICET Argentina.

\end{document}